\documentclass[journal]{IEEEtran}
\ifCLASSINFOpdf
\else
\fi
\usepackage{graphicx}
\usepackage{hyperref}
\usepackage{indentfirst} 
\usepackage{multirow} 
\usepackage{booktabs}
\usepackage{makecell}
\usepackage{url}
\usepackage{cite}
\usepackage{graphicx}
\usepackage{amsmath}   
\usepackage{color}
\usepackage{amssymb}
\usepackage{subfigure}

\usepackage{multirow}
\usepackage{fancyhdr}
\usepackage{epic}
\usepackage{eepic}
\usepackage{amsmath}
\usepackage{threeparttable}
\usepackage{amscd}
\usepackage{here}
\usepackage{graphicx}
\usepackage{lscape}
\usepackage{tabularx}
\usepackage{longtable}
\usepackage{comment}
\usepackage{amsmath,amssymb,amsfonts}
\usepackage{graphicx}
\usepackage{textcomp}
\usepackage{booktabs}
\usepackage{float}
\usepackage{algorithm}  
\usepackage{algorithmicx}  
\usepackage{algpseudocode}
\usepackage{extarrows}
\usepackage{subfigure}

\usepackage{lipsum} 

\usepackage{tikz,xcolor,hyperref}

\definecolor{lime}{HTML}{A6CE39}
\DeclareRobustCommand{\orcidicon}{%
	\begin{tikzpicture}
		\draw[lime, fill=lime] (0,0) 
		circle [radius=0.16] 
		node[white] {{\fontfamily{qag}\selectfont \tiny ID}};
		\draw[white, fill=white] (-0.0625,0.095) 
		circle [radius=0.007];
	\end{tikzpicture}
	\hspace{-2mm}
}

\foreach \x in {A, ..., Z}{%
	\expandafter\xdef\csname orcid\x\endcsname{\noexpand\href{https://orcid.org/\csname orcidauthor\x\endcsname}{\noexpand\orcidicon}}
}

\begin{document}
%
\title{Fabric-SCF: A Blockchain-based Secure Storage and Access Control Scheme for Supply Chain Finance}
%
%
%

\author{
	Dun Li\orcidA{}, 
	Dezhi Han\orcidB{}, 
	Noel Crespi,
	Roberto Minerva\orcidR{},
	Zhijie Sun\orcidE{}
	
	\thanks{
		Dun Li is with the College of Information Engineering at Shanghai Maritime University, China and Telecom SudParis, IMT, Institut Polytechnique de Paris, France. E-mail:lidunshmtu@outlook.com.
		(Corresponding Author)
	}  
	\thanks{
		Dezhi Han is with the College of Information Engineering at Shanghai Maritime University, China. E-mail:dzhan@shmtu.edu.cn
		(Corresponding Author)
	}
	\thanks{
		Noel Crespi and Roberto Minerva are with Telecom SudParis 9 Rue Charles Fourier, 91000 Evry, France.
		Email: noel.crespi@it-sudparis.eu
	}
	\thanks{
		Zhijie Sun is with the College of Information Engineering at Shanghai Maritime University, China. E-mail: S19117157825@163.com.
	}

} 	

%
%

\markboth{Journal of \LaTeX\ Class Files,~Vol.~14, No.~8, August~2015}%
{Shell \MakeLowercase{\textit{et al.}}: Bare Demo of IEEEtran.cls for IEEE Journals}

\maketitle

\begin{abstract}
	Supply chain finance(SCF) is committed to providing credit for small and medium-sized enterprises(SMEs) with low credit lines and small financing scales. 
	The resulting financial credit data and related business transaction data are highly confidential and private.
	However, traditional SCF management schemes mostly use third-party platforms and centralized designs, which cannot achieve highly reliable secure storage and fine-grained access control.
	To fill this gap, this paper designs and implements Fabric-SCF, a secure storage and access control system based on blockchain and attribute-based access control (\textbf{ABAC}) model.
	This scheme uses distributed consensus to realize data security, traceability, and immutability.
	We also use smart contracts to define system processes and access policies to ensure the efficient operation of the system.
	To verify the performance of Fabric-SCF, we designed two sets of simulation experiments.
	The results show that Fabric-SCF achieves dynamic and fine-grained access control while maintaining high throughput in a simulated real-world operating scenario.
\end{abstract}


\IEEEpeerreviewmaketitle

\section{Introduction}
\label{sec:Introduction}
\IEEEPARstart{S}{upply} 
chain finance(\textbf{SCF}) enables enterprises of different sizes to have better opportunities for joint development.
Under the situation of economic globalization, small and medium-sized enterprises(\textbf{SMEs}) have gradually become an important driving force for national economic development \cite{Zulqurnain2020DoesSC}. 
SMEs need sufficient cash flow for their operations, but capital inflows and expenditures for supply chain operations occur at different times, so access to financing services is often required to cover shortfalls in capital flow.
However, traditional financial services treat SMEs as isolated entities \cite{Yue2019AGI}, ignoring their business potential in the supply chain, making them face long-term financing difficulties\cite{Somjai2019GoverningRO}.
As a result, SCF plays a significant role in serving the entity economy and SMEs. 
SCF combines capital operations with the supply chain and establishes a sound transaction structure \cite{Gelsomino2016SupplyCF}.
SCF has entered the information and intelligent stage \cite{Caniato2019SupplyCF}. 
Through the construction of enterprise information systems, the difficulties of information asymmetry among supply chain stakeholders have been alleviated, the cost of information circulation has been reduced, and convenient and efficient financial services can be provided \cite{Guo2020InfluencesOS}.

However, SCF involves large amounts of digital data.
The flow of productions, information, and capital is often seen as isolated across functions and parties.
In traditional SCF management solutions, the initiation of business processes depends on sequential input and manual validation.
The high cost and complexity of these solutions, security flaws, and time-consuming processing have caused the development of SCF to encounter bottlenecks\cite{Omran2017blockchaindrivenSC}.
In addition, in the financing business of SCF, the credit transmission capacity of core enterprises is limited.
And SMEs have problems such as irregular financial statements and inadequate management systems.
Therefore, it is difficult for financial institutions to effectively control risks through the information provided, and it is urgent to add effective privacy protection mechanisms and access control mechanisms to the SCF management plan to help financial institutions obtain information efficiently and securely \cite{Li2020FabricChainC}.

Interestingly, blockchain technology can solve the current problems of information asymmetry, lack of visibility in the transaction process, and possible joint fraud in the core business model.
As the underlying technological framework for cryptocurrencies, blockchain is decentralized and immutable \cite{Zheng2017AnOO, Li2020ASO}.
Blockchain is a transparent, secure decentralized system based on an immutable model, so it is seen as a solution to change the rules in the traditional supply chain industry. 
Currently, blockchain projects can be divided into four main categories which are cryptocurrency, platform, application, and asset tokenization.
Mainstream blockchain platforms have been widely used in  healthcare \cite{Omar2019PrivacyfriendlyPF, Tanwar2020blockchainbasedEH, Khatoon2020ABS, Dimitrov2019blockchainAF, Bhattacharya2020BinDaaSBD}, 
smart cities \cite{Bhushan2020blockchainFS, Sharma2018blockchainBH, Aujla2020BlockSDNBF}, energy networks \cite{Pieroni2018SmarterCS, Munsing2017blockchainsFD, Li2019PanoramicIM, Wang2019EnergyCA}, 
Internet of Things (IoT) \cite{Dorri2019LSBAL, Ding2019ANA, Lao2020GPBFTAL, Liu2020FabriciotAB}, 
Internet of Vehicles (IoV) \cite{Kang2019TowardSB, Jiang2019blockchainBasedIO, Javaid2020ASP, Wang2019AnIA, Liu2021BehaviorAA, Zhou2019blockchainAC, Liu2020blockchainBT}, and education\cite{Turkanovi2018EduCTXAB, Harthy2019TheUB, Oganda2020blockchainES, Li2021MOOCsChainAB}.
Thus, a blockchain-based SCF solution is urgently needed and feasible.

Besides, with the emergence of the smart contract-enabled blockchain projects such as Ethereum and Hyperledger, the blockchain has entered the 2.0 era.
Smart contracts are programmable and Turing-complete \cite{Wang2019blockchainEnabledSC}. 
The transaction can be initiated automatically according to the rules set by the code.
The introduction of smart contracts is expected to change the existing transaction mode of SCF and reconstruct the breakthrough technology of society from the underlying infrastructure.
Moreover, in a realistic access control strategy, it is necessary to approve or deny the user's request based on the environmental conditions implemented by the current policy. 
In the open network environment of blockchain, the attribute-based access control (\textbf{ABAC}) model is suitable and efficient as a flexible and fine-grained access control method \cite{Li2017FlexibleAF}.
This model mainly determines whether the data requester has the correct attributes to determine the data requester's access control rights to private data resources.

So far, scholars have given extensive attention to issues such as risk pricing, profit enhancement, and business models for SCF.
However, there have been few attempts to apply blockchain to SCF management systems for optimization and improvement.
In this context, we propose a blockchain-based SCF management system that digitizes the business processes in SCF, optimizes them by deploying smart contracts, and stores them in the blockchain in the form of data streams, ensuring privacy, immutability, and traceability.
We also introduce the ABAC model for access control to ensure efficient and secure access.

Specifically, the main contributions of this paper can be summarized as follows.  
\begin{itemize}
	\item{We apply blockchain to SCF management systems, enabling decentralized management, secure storage, and multi-party maintenance of financing projects with the help of distributed consensus and authentication mechanisms.}
	\item{Based on the basic framework of blockchain, this paper designs an auxiliary architecture based on ABAC to achieve fine-grained access control.}
	\item{In this paper, we use smart contracts to define multi-tier data structures, access policies, and system workflow to ensure high-speed data storage, retrieval, query, and fine-grained access control.}
	\item{We have completed the preliminary development of Fabric-SCF and proved its performance through simulation experiments.}
\end{itemize}

The rest of this paper is organized as follows.  
We first introduce the related work in Section \ref{sec:Related Work}. 
Then, in Section \ref{sec:preliminaries}, we present the relevant preliminaries. 
Next, we present the system models, assumptions, and design goals in Section \ref{sec:System model and design}. 
In Section \ref{sec:Experiment and Comparison}, 
the performance of the proposed model is analyzed based on experimental results. Finally, the conclusions and future work are introduced in Section \ref{sec:Conclusion and Further Work}.

\section{Related Work}
\label{sec:Related Work}
In this section, we first survey the related works about the SCF model in Section \ref{sec:Supply chain financing model} and then present the application of blockchain and smart Contract in SCF scenarios in Section \ref{sec:Applications of blockchain in SCF} and Section \ref{sec:Applications of Smart Contracts in SCF} respectively. 

\subsection{SCF Model}
\label{sec:Supply chain financing model}
SCF has long been recognized as an important intersection between the fields of trade finance and supply chain management.
The empirical results show that, a well-structured SCF solution can significantly enhance the efficiency of capital flows for SMEs \cite{Wang2020DriversAO,Pan2020TheIO,Pellegrino2019SupplyCF,Lin2017SelectingTS,Lam2021TheIO,Ali2019PredictingSC,Garg2021ModelingTS, Nguema2021TheEO,Gelsomino2019AnOS, Ding2020FinancingAC}.

Many researchers have worked to build new models of SCF that improve the efficiency of data flow and privacy protection mechanisms.
To hedge financial risk, traditional SCF relies on credit rating \cite{Moretto2019SupplyCF}.
In \cite{Su2015SimulationOG,Su2016TheCR}, authors used multi-agent simulation technology instead of absolute rationality to design a SCF simulation model based on Simon's bounded rationality.
Lu \textit{et al}. \cite{Lu2019SupplyCF} evaluated the bank's fair pricing rate under the guarantee model and found that the financial model of third-party guarantees may not always be applicable.
Xiao \textit{et al}. \cite{Xiao2019ResearchOS} and Liao \textit{et al}. \cite{Liao2019INTEGRATINGBA} try to use entropy weight, analytic hierarchy process, multi-expert decision model, and other methods to sort suppliers to improve the efficiency of the model.
In \cite{Song2021BigDA}, big data analytics are applied to build business processing platforms for SCF and structure business data.
The review of Jia \textit{et al}. \cite{Jia2020SustainableSC} summarized that sustainable supply chain finance (SSCF) can significantly enhance the sustainability and development performance of SMEs.
The work of \cite{AbdelBasset2020AND,Tseng2019ImprovingTB, Imronudin2020SustainableSC, Tseng2018DecisionmakingMF, Bui2020SustainableSC} confirms this view.
Emtehani \textit{et al}. \cite{Emtehani2021AnOI} proposed an effective SCF decision framework from an optimization business perspective that assists in operational financing decisions and improves financial efficiency from an algorithmic perspective.
Then, Sang \cite{Sang2021ApplicationOG} proposed a risk assessment strategy based on genetic algorithm and neural network method to reduce the probability of profit damage of the fund providers in SCF.
Yin \textit{et al}. \cite{2021Edge} proposed an edge-intelligent SCF model based on B2B platform, which has obvious efficiency advantages in responding to the changing trend of the parameters related to the SCF model of B2B platform.
Abbasi \textit{et al}. \cite{Abbasi2019ResearchOM} designed an IoT-based SCF model and proves that the credit risk measurement model is credible.

The above models incorporate many new techniques in operations research, machine learning, neural networks, IoT, and edge computing to try to improve the efficiency of SCF business processes and enhance data security, but they all need to rely on third-party platforms.

\subsection{Applications of blockchain in SCF}
\label{sec:Applications of blockchain in SCF}
The applications of blockchain technology originally originated from cryptocurrencies, as the underlying database and maintenance technology.
Due to decentralization, consensus mechanism, and unalterability, blockchain technology has broad industrial application prospects.

In \cite{Liu2021LiteratureRO, Tribis2018SupplyCM}, authors reviewed the feasibility of applying blockchain technology in SCF and confirmed that the technical characteristics of blockchain can effectively promote the development of SCF.
Specifically, Du \textit{et al}. \cite{Du2020SupplyCF} proposed a blockchain-based business management platform for SCF, which solves the distrust problem among supply chain participants.
Liu \textit{et al}. \cite{Liu2021AHB} proposed a hybrid chain model combining a public chain-based consistency algorithm and a federated chain-based consistency algorithm to process each account's transactions in parallel, serving the supply chain financial data management of engineering projects and enabling more efficient transaction authenticity auditing, risk assessment, and credit delivery for core enterprises.
Li \textit{et al}. \cite{Li2020blockchaindrivenSC} proposed a solution for SCF and showed the operation process of three SCF models on the platform.
Wag \textit{et al}. \cite{Wang2021ApplicationOB} proposes a model that combines blockchain technology and SCF gaming and uses blockchain to connect banks, other financial institutions, and core companies.
Chen \textit{et al}. \cite{Chen2020ABS} proposed BCAutoSCF, a blockchain-powered SCF platform dedicated to serving the automotive retail industry, enabling reliable, transparent, and traceable business inquiries and data access.
Similarly, in \cite{Wang2020ResearchOT, Huang2018ResearchOT, Jiang2020ApplicationOB, Wang2021TheEO, Li2020TheRO, Yuyan2020TheRO, Yaksick2019OvercomingSC, Dong2021blockchainEnabledDS, Xu2020ResearchOA, Zhang2021AnalysisOT, FernndezCarams2019TowardsAA}, blockchain technology has been used to enhance the privacy of business management processes for SCF.

\subsection{Applications of Smart Contracts in SCF}
\label{sec:Applications of Smart Contracts in SCF}
Notably, smart contracts have Turing-complete programmability and can be used to deploy the specific implementation of SCF-related business.
Kim \textit{et al}. \cite{Kim2018TowardsAO} proposed an ethereum-based blockchain platform and transformed business processes into smart contracts to ensure data traceability.
Also based on Etherium, Huertas \textit{et al}. \cite{Huertas2018EximchainSC} proposed an intelligent contract-based ecosystem allowing SMEs to quickly implement bespoke SCF solutions called Eximchain, which uses a consensus protocol and a secondary voting-based governance model to provide practical, time-limited security guarantees.
Ref. \cite{Terzi2019TransformingTS, Liu2020ANS, Su2018SmartsupplySC} also proposed similar SCF solutions based on smart contracts deployed in Etherium.
As in this paper, Ma \textit{et al}. \cite{Ma2019ThePP} proposed a Hyperledger Fabric-based solution that adapts to a wide variety of scenarios and complex business processes using segmented permissions, privacy protection mechanisms, etc.
In \cite{Schr2020DecentralizedFO}, more financial models based on smart contracts are also discussed.

In summary, these models attempt to build decentralized blockchain-based models from a privacy-preserving perspective, using smart contracts to implement business processes in SCF.
However, most of these models have been tested on public chains, and there is a lack of industrial-grade platforms that can be used in practice.

\section{Preliminaries}
\label{sec:preliminaries}

This section describes the main business processes for SCF.
Tab. \ref{tab: Key Notations} presents the notations used in this paper.

\begin{table}[htbp]
	\caption{Key Notations}
	\label{tab: Key Notations}
	\centering
	\begin{tabular}{p{75pt}| p{130pt}}
		\toprule
		\hline
		\textbf{Notation} & \textbf{Description} \\
		\hline
		$CE$ & Core enterprise \\
		$FI_i$ & Financial institution $i$ \\
		$FE_i$ & Financing enterprise $i$ \\
		$Sp_i$, $Dt_j$ & Supplier $i$, Distributor $j$ \\
		$CF_i$ & Credit facility $i$ \\
		$Col_i$ & Collateral $i$ \\
		$P_i-_j$ & Productons $i-j$ \\
		$ARD_i$ & Accounts receivable document $i$ \\
		$PC_i$ & Prepayment Contract $i$ \\	
		$FO_i$ & Financing order $i$ \\
		$FP_i$ & Financing Project $i$ \\
		$CA$ & Certificate Authority \\
		$Ct_i$ & Certificate $i$ \\
		$Cf_i$ & Config file of $i$ \\
		$pubk_i$ & The public key of $i$ \\
		$prik_i$ & The private key of $i$ \\
		$\sigma_{i}$& The anonymous credential of entity $i$ \\
		$RC_{(x)}$ & The storage credential of content $x$ \\
		$CT_{(x)} $ & Encrypted content for $x$\\
		$HV_{(x)}$& The hash value of $x$ \\
		$ST_{(x)}$& The encrypted content of $x$ \\
		\hline
		\bottomrule
	\end{tabular}
\end{table}

\begin{figure}[H]
	\centering
	\includegraphics[width=\linewidth]{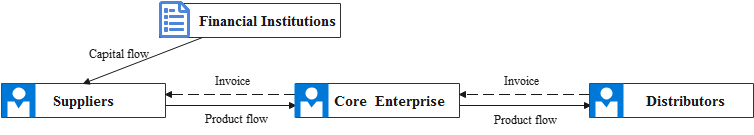}
	\caption{An example of SCF framework.}
	\label{fig:SCF-example}
\end{figure}

A typical SCF framework consists of a core company, suppliers, distributors, and financial institutions.
Fig.\ref{fig:SCF-example} shows an example.

SCF uses self-repaying trade finance based on the core enterprise.
Compared with traditional credit business, SCF focuses more on the real trade background between the upstream and downstream of the supply chain. 
In the past, commercial bank examination, and approval mainly on credit enterprises due diligence, a full range of examination and approval of its credit level, whether there is real estate can be mortgaged, cash flow is stable to ensure the future ability to pay the debt, although lending will also be entrusted to pay to ensure the use of funds, less attention to the overall supply chain.
Based on different financing directions and mortgage targets, there are three main types of SCF businesses, namely accounts receivable financing, inventory financing, and prepayment financing.

\subsection{Accounts Receivable Financing}
\label{sec:Accounts Receivable Financing}
The accounts receivable financing model is a financing business provided by a firm based on accounts receivable arising from a real trade between a seller and a buyer, which is repaid by the contracted accounts receivable.
Specifically, the supplier $Sp_i$ first enters into a transaction with the core enterprise $CE_i$, which is downstream in the supply chain as  $Sp \stackrel{P_i-_j}{\longrightarrow} CE$, and the core enterprise $CE_i$ issues an accounts receivable document to the supplier $Sp_i$ as $CE \stackrel{ARD}{\longrightarrow} Sp$.
Then, the supplier $Sp_i$ assigns the receivables document $ARD_i$ to the financial institution $FI_i$ and the downstream supplier in the supply chain makes a payment commitment to the financial institution $FI_i$ as $Sp \stackrel{ARD}{\longrightarrow} FI$.
The financial institution $FI_i$ provides a credit facility $CF_i$ to the supplier $Sp_i$ as $FI \stackrel{CF}{\longrightarrow} Sp$. 

\subsection{Inventory Financing}
\label{sec:Inventory Financing}
Inventory financing uses the productions $P_i-_j$ in the trade process as the collateral $Col_i$ for financing. Generally, it occurs when the supply of $Sp_i$ has a large inventory or slow inventory turnover, which leads to high pressure on capital turnover and enterprises use existing products. $P_i-_j$ get cash in advance as $Sp \stackrel{P_i-_j}{\longrightarrow} FI$.
The financial institution $FI_i$ can issue the corresponding credit facility $CF_i$ based on the evaluation of the third-party regulatory company as $FI \stackrel{CF}{\longrightarrow} Sp$.
If the SME loses the ability to repay during the repayment period, the financial institution $FI_i$ can find the core enterprise $CE$, sign a purchase agreement with the core enterprise $CE$ as $FI \stackrel{P_i-_j}{\longrightarrow} CE$, and complete the repayment as $CE \stackrel{Cash}{\longrightarrow} FI$.

\subsection{Prepayment Financing}
\label{sec:Prepayment Financing}
Based on inventory financing, prepayment financing has been developed. 
The buyer $Br_i$ negotiates full payment to the seller $Sr_i$ on the condition that a percentage of the deposit is paid.
The seller $Sr_i$ ships the productions $P_i-_j$ in accordance with the purchase and sale contract and the cooperation agreement, and the productions $P_i-_j$ arrive with a pledge as security $Col_i$ for the financial institution $Fi_i$.
In the specific process, the core enterprise $CE_i$ provides a purchase and sale contract $PC_i$ to the distributor $Dt_i$.
Then the distributor $Dt_i$ acts as a financing enterprise $FE_i$ to get credit facility as $CE \stackrel{P_i-_j}{\longrightarrow} FI$ and $Fi_i \stackrel{CF}{\longrightarrow} Dt_i$.
Distributor $Dt_i$ submits the delivery deposit to the financial institution $Fi_i$ in installments, and the financial institution $Fi_i$ then informs the core enterprise in installments to deliver the productions $P_i-_j$ to the distributor $Dt_i$.

\section{System Model and Design}
\label{sec:System model and design}
This section introduces the architecture of Fabric-SCF. 
In Section \ref{sec:System Architecture}, we present the designed system architecture. 
Section \ref{sec:Data Structure} introduces the defined data structure. 
Section \ref{sec:Smart Contract Design} shows the details of the smart contract design. 
In Section \ref{sec:Workflow}, we present the workflow of Fabric-SCF.

\subsection{System Architecture}
\label{sec:System Architecture}

\begin{figure}[H]
	\centering
	\includegraphics[width=\linewidth]{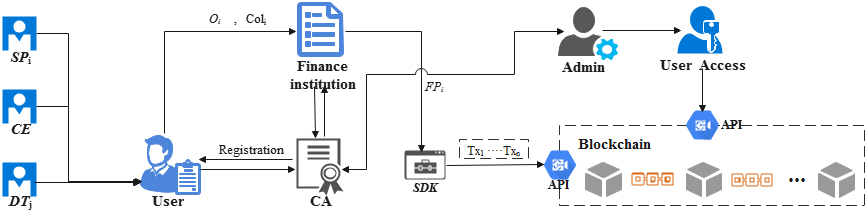}
	\caption{Architecture of Fabric-SCF.}
	\label{fig:System Architecture}
\end{figure}

The proposed Fabric-SCF is composed of consortium blockchain, user, admin, and user access to form the underlying structure as shown in Fig. \ref{fig:System Architecture}.

\textbf{Blockchain} is a distributed network composed of trusted nodes for data synchronization and storage.   
In this network, authentication is realized through digital certificates to ensure the security and integrity of data in the system.

\textbf{User} contains business participants. Logically, business participants contain mainly core enterprise $CE$, supplier $SP_i$, and Distributor $Dt_j$.

\textbf{Admin} is responsible for managing the access to blockchain system.
After entering the blockchain system through the validation of \textit{CA}, it needs to develop access policies, distribute user access to data, and maintain access policies.

\textbf{User Access} is a collection of access commands.   
In this network, Fabric-SCF allows all parties involved in SCF to be authorized to send data flows, access data, and perform $add()$, $query()$, $update()$ functions.

The three types of business correspond to different stages of the supply chain business.
All credit facility $CF_i$ provided by financial institutions $FI_i$ is based on the stage of business with the core enterprise $CE$.
The Fabric-SCF system uses the blockchain as a database and is divided into a supply-side and a distribution-side in terms of business processes with the core enterprise as the center.
Sellers $Sr_i$ and buyer $Br_j$ can be unified to supplier $Sp_i$ and distributor $Dt_j$. 
For financing business, supplier $Sp_i$ and distributor $Dt_j$ can be grouped as financing companies $FE_i$ or $FE_j$ as users.
The financial institution $FI_i$ uses inventory $P_i-_j$, accounts receivable documents $ARD_i$ or prepayment contracts $PC_i$ as collateral $Col_i$ and the financing order $FO_i$ to lend credit facility $CF_i$ to the financing enterprise $FE_i$ based on the business transactions between the financing enterprise $FE_i$ and the core enterprise $CE$.
All users and financial institutions $FI$ need to be authorized by \textit{CA}, and all data access requests need to go through \textit{Admin}.
The data flow is controlled by smart contracts, and the corresponding data is stored in the ledger in the blockchain, thus forming a closed loop of data exchange.

\subsection{Data Structure}
\label{sec:Data Structure}
\begin{figure}[htbp]
	\centering
	\includegraphics[width=\linewidth]{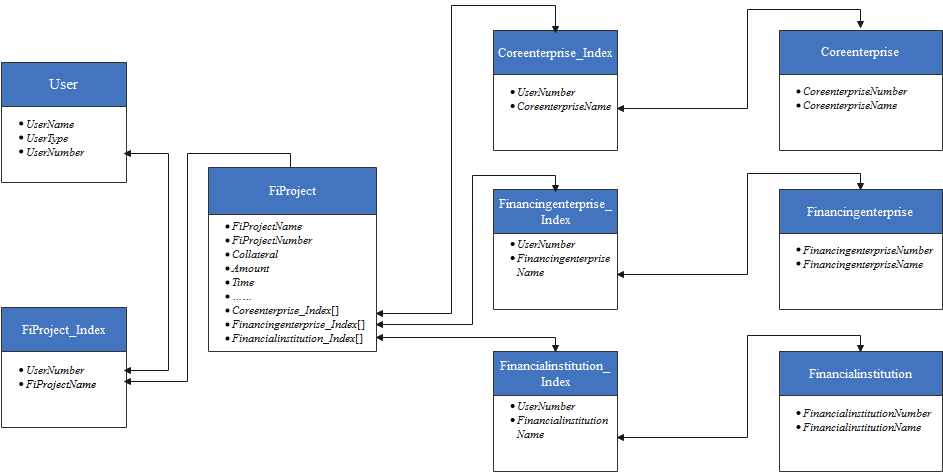}
	\caption{Relationship between data structures.}
	\label{fig:Data Struct}
\end{figure}

Blockchain can be summarized as a distributed state machine. All nodes start from the same creation state, run the transactions in the block that have reached a consensus in turn, and drive the states of each node to change according to the same operation sequence (add, delete, update, and query).
For all nodes in the implementation of the same number of blocks after the transaction, the state is completely consistent.
To ensure that the world state can be updated quickly when the transaction is executed, the design and implementation of the world state scheme should consider the fast search and efficient update of the state data.
To ensure system efficiency and reduce system latency, nine data structures are designed in this paper.
The index relationship is shown in Fig. \ref{fig:Data Struct}.

Defining the data structures in this way facilitates the design of uniform data access operations and reduces the complexity of the system.
The multiplicity of mappings between data structures is shown in Eq. \ref{eq:index}.

\begin{small}
	\begin{equation}
		\label{eq:index}
		\begin{cases}
			User\rightarrow FC_\_Index &= 1:n \\
			FC\_Index \rightarrow FiProject &= 1:1 \\
			FC \rightarrow CE\_Index &= 1:1 \\
			FC \rightarrow FE\_Index &= 1:1 \\
			FC \rightarrow Fi\_Index &= 1:1 \\
			CE\_Index \rightarrow CE &= 1:1 \\
			FE\_Index \rightarrow FE &= 1:1 \\
			FI\_Index \rightarrow FI &= 1:1 \\
		\end{cases}
	\end{equation}
\end{small}

\subsubsection{User}
The data structure of $User$ corresponds to the system role of $User$, and it mainly contains three parts.
Where $UserName$ directly reflects the user's name, and $UserType$ corresponds to the main participants of the system process $SP_i$, $Dt_j$, $CE$.
$UserNumber$ is calculated by Eq. \ref{eq:UserNumber}.

\begin{equation}
	\label{eq:UserNumber}
	UserNumber_i = HV(UserName_i, pubKey_i)
\end{equation}

\subsubsection{FiProject}
As the core data block, it contains many attributes. $FiProject$  has a one-to-one correspondence with $FiProject\_Index$.
The core segments are as follows.

i) $Collateral$: The collateral $Col_i$ of the financing enterprise $FE_i$ to the financial institution $FI_i$ when handling the financing business includes inventory $P_i-_j$, accounts receivable document $ARD_i$, and prepayment $Pm_i$ as Eq. \ref{eq:flag}.

\begin{equation}
	\label{eq:flag}
	Col_i=\begin{cases}
		\vspace{1ex}	
		P_i-_j \\   \vspace{1ex}
		ARD_i \\  
		Pm_i
	\end{cases}
\end{equation}

ii) $Amout$: The interest rate and amount of the credit facility $CF_i$ specified in the financing project.

iii) $Time$: The timeline of the credit facility $CF_i$ in the financing project $FP_i$.

iv) $Index$: The index of the core enterprise CE, Financing enterprise $FI_i$ and financial institution $FI_i$ related to the financing project $FP_i$.

\subsection{Smart Contract Design}
\label{sec:Smart Contract Design}

Smart contract is the key to realizing the business logic and access control of the system.
To maximize the protection of data privacy and integrity while ensuring the efficiency of the system, this paper mainly designs the smart contract from the perspective of user management, financing project management, and access control.

\subsubsection{User Management Contract}
\label{secUser Management Contract}
In our proposed system, the management of users mainly involves creation and query.
To ensure the uniqueness of the identity of participants of multiple organizations, it is stored in the state database of the blockchain as Fig. \ref{fig:User} shows.

i) $createUser()$:
$User$ registration is the first step for to participate in the system, and $createUser()$ needs to generate a $UserNumber$ based on the $UserName$ and $UserType$ as a unique identifier.

ii) $queryUser()$:
Thin function can query the $user$ information based on $UserName$ and $UserNumber$. In addition, it is also needed to check if the user already exists before the member is registered.

iii) $checkUser()$:
If $user$ wants to be authorized by an access policy, it needs to pass verification first to verify the $Ct$, $SingnID$, $Pubk$, $Prik$ and $UserType$.

\begin{figure}[H]
	\centering
	\includegraphics[width=\linewidth]{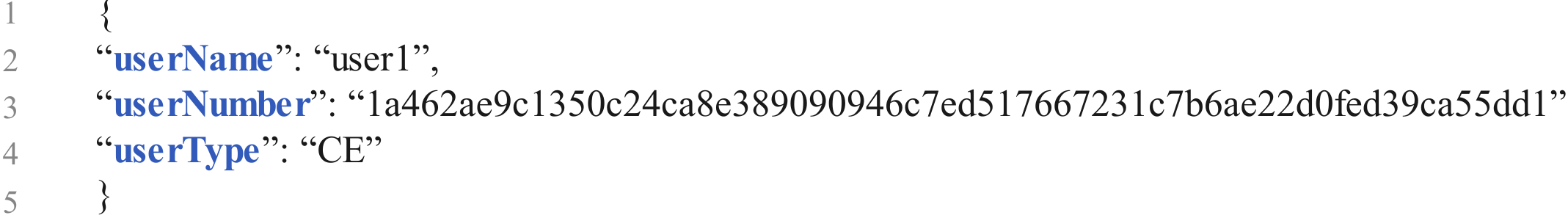}
	\caption{An example of member data.}
	\label{fig:Usere}
\end{figure}

\subsubsection{Financing Project Management Contract}
\label{secFinancing}
According to user requests, related financing project data access operations mainly include $addFiproject()$, $queryFiproject()$, $updateFiproject()$, and $deleteFiproject()$. 
Besides, we also define $checkFiproject()$ and $checkUdvInfo()$ method to ensure the legitimacy and reliability.
These operations correspond to different entities, different authorizations, and different processes.

i) $addFiproject()$: 
When a user submits a request to add a contract, $checkFiproject()$ will be triggered first to check the reasonableness of the contract information. If the contract information is reasonable, the contract will be published and the contract and all operations on the contract will be packaged and written to the blockchain ledger.
The adding and judgment logic is shown in \textbf{Algorithm}. \ref{al:addproject}.

\begin{algorithm}[h]
	\caption{addFiProject()}
	\label{al:addproject}
	\begin{algorithmic}[1]
		\Require 
		FiProject(FP)
		\Ensure
		Ok or Error
		\State @implement SmartContract Interface;
		\State APIstub ChaincodeStub ← Invoke();
		\If{CheckFiProject(FP) == False}
		\State \Return{Error(BadFP)};
		\EndIf
		\State Id ← Sha256(FP.FiProjectName + FPFiProjectNumber);
		\State err ← APIstub.PutState(Id, FP);
		\If{err! = null}
		\State \Return{Error(err.Error())};
		\EndIf
		\State \Return{Ok};
	\end{algorithmic}
\end{algorithm}

ii) $queryFiproject()$: 
The query function is very common and all users can implement queries for details related to the authorized contract data. Both the financing project name $FiProjectName$ and financing project number $FiProjectNumber$ can be used as the basis for queries.

iii) $updateFiproject()$: 
This function updates the order information. 
In some special cases, the contract details such as contract term, interest rate due, etc. need to be changed.
Similar to $addFiproject()$, the update process first triggers $checkUdvInfo()$ to check the reasonableness of the updated information and posts the verified update information, and updates the ledger.

iv) $deleteFiproject()$: 
Deleting a contract usually happens by mistake, or when a contract is re-entered.
When the user issues a request to delete an order, $checkFiproject()$ is first triggered to check the reasonableness of the contract information, and then $queryFiproject()$ is called to check the status of the order.

v) $checkFiproject()$: 
When the user submits $addFiproject()$, this method is used to check whether the key fields of the financing project information are reasonable.

vi) $checkUdvInfo()$: 
This method is used to verify the legitimacy of the update related to $updateFiproject()$.

\subsubsection{Access control contract}
\label{AccessC}
The access control management contract is used to verify that a user's data access request satisfies the access control policy set by the administrator and consists mainly of $Auth()$, $GetAttrs()$, $CheckPolicy()$ and $CheckAccess()$.

i) $Auth()$: 
The public key distributed to the user by \textit{Admin} can be used to encrypt the request.
Based on this, this method is used to authenticate the user request and further determine the user's identity.

ii) $GetAttrs()$:
The attribute fields contained in the user's request are resolved by calling the $GetAttrs()$ method after verifying the user's identity.
The attribute fields contained in the request are the subject attribute and the object attribute $\{S, O\}$.
After further determining the environment attribute $E$, the attribute combination can be obtained as $\{S, O, E\}$.

iii) $CheckAccess()$:
This method is the core of the implementation of access control management.
First, the ${S, O}$ attribute is obtained via the $GetAttrs()$ method, then the $QueryPolicy()$ method is called using the former to obtain the corresponding ABACPolicy. 
If the method returns null, then no policy supports the request and the request is invalid. 
Then, the request is verified by validating the policy that matches, and if the attributes $E$ and $A$ in the policy are both satisfied, the request passes the verification. 
The pseudo-code of the $CheckAccess()$ is shown in \textbf{Algorithm}. \ref{al:CheckAccess()}.

\begin{algorithm}[h]
	\caption{CheckAccess()}
	\label{al:CheckAccess()}
	\begin{algorithmic}[1]
		\Require 
		ABAC\_Request
		\Ensure
		OK or Error
		\State \{$S_u$, $O_u$, $E_u$\} ← GetAttrs(ABAC\_Request);
		\State Policy =\{$Policy_1$,...$Policy_n$\} ← $QueryPolicy(S_u, O_u)$;
		\If{Policy == Null}
		\State \Return{Error()};
		\EndIf
		\For {Policy in \{$Policy_1$,...$Policy_n$\}}
		\State \{. . . , $P_p$, $E_p$\} ← Policy
		\If{Value(Pp) == 1 \&\& Eu $\cap$ Ep !=Null} 
		\State \Return{OK};
		\EndIf
		\EndFor
		\State \Return{Error()};
	\end{algorithmic}
\end{algorithm}.

\subsubsection{Policy Contract}
The PC provides the following methods to operate ABACPolicy.

i) $AddPolicy()$: The PSC needs to run the $CheckPolicy()$ method before calling this method to pass it to add the policy, and the policy can only be written to the SDB and the blockchain after the policy is legal. 
The details are shown in \textbf{Algorithm}. \ref{al:PSC.AddPolicy()}.

\begin{algorithm}[h]
	\caption{AddPolicy()}
	\label{al:PSC.AddPolicy()}
	\begin{algorithmic}[1]
		\Require 
		ABACP
		\Ensure
		OK or Error
		\State @implement SmartContract Interface;
		\State APIstub ChaincodeStub ← Invoke();
		\If{err! = null}
		\State \Return{Error(BadPolicy)};
		\EndIf
		\State Id ← Sha256(ABACP.S + ABACP.O);
		\State err ← APIstub.PutState(Id, ABACP);
		\If{err! = null} 
		\State \Return{Error(err.Error())};
		\EndIf
		\State \Return{OK};
	\end{algorithmic}
\end{algorithm}.

ii) $DeletePolicy()$: This method will be called in two ways. Firstly, the administrator will actively call this method to delete an ABACPolicy.
Secondly, if a policy is found to have expired during the execution of the $CheckAccess()$ method, then this method will be called automatically to delete the useless policy. 
%

iii) $UpdatePolicy()$: This method will be called in two ways. Firstly, the administrator will actively call this method to delete an ABACPolicy.
This method is called when the administrator needs to modify an ABACPolicy. When calling this method, you need to write a modified record into the SDB and blockchain. 
The method has finally executed the $addPolicy()$ method after completing the policy update and re-increasing the modified policy into the blockchain.

iv) $QueryPolicy()$: All strategies are stored in Status Database CouchDB, and the administrator can query the details of the ABACPolicy by attribute $S$ or $O$.

\subsection{Workflow}
\label{sec:Workflow}

\begin{figure}[ht]
	\centering
	\includegraphics[width=\linewidth]{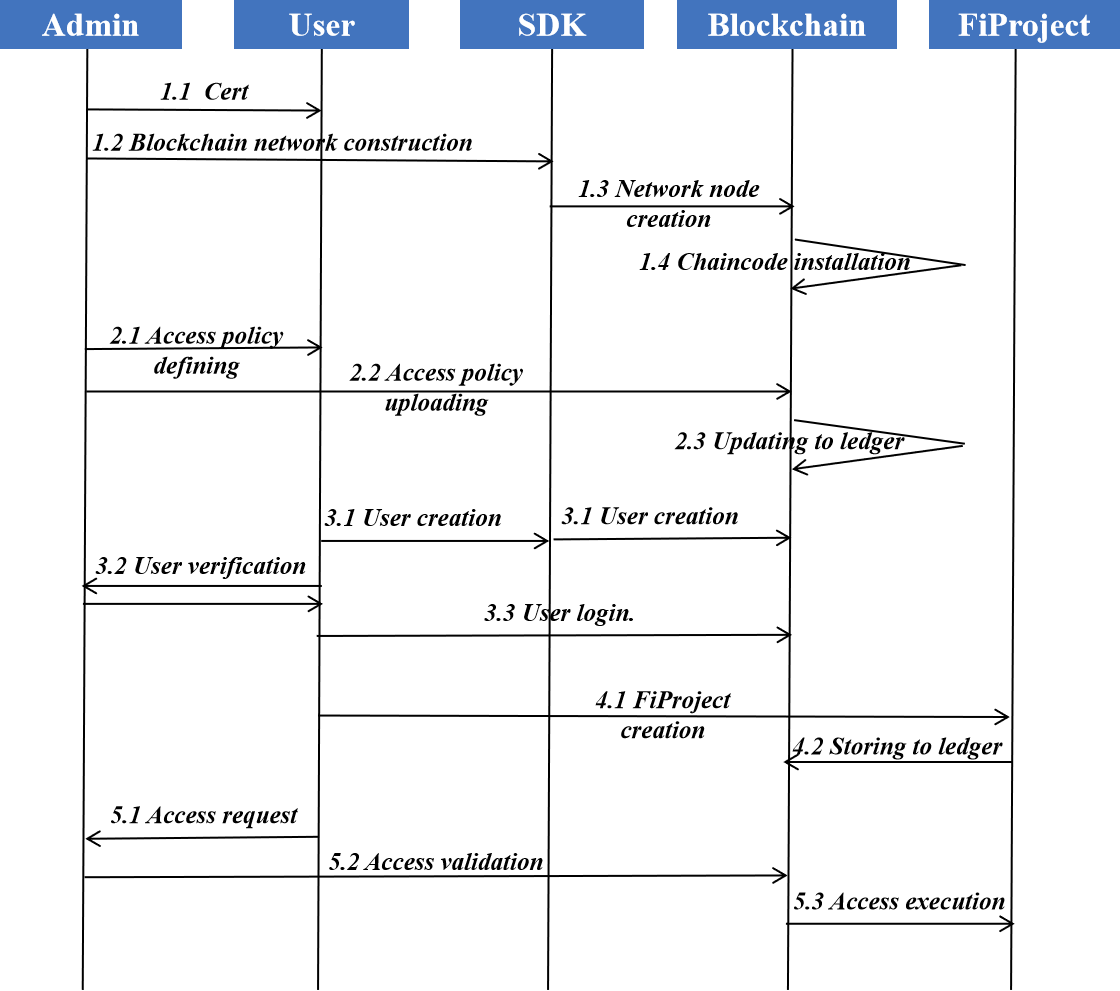}
	\caption{Workflow}
	\label{fig:W}
\end{figure}

The system workflow consists of four parts.
1) Blockchain network initialization.
2) Access policy deployment.
3) User registration.
4) Financial project management.
5) User Access Assignment
We present the details of the five parts in Fig \ref{fig:W}.

\section{Experiment and Comparison}
\label{sec:Experiment and Comparison}
This section describes the experimental procedure and the comparison of results to show the functionality and performance of the proposed Fabric-SCF.
Section \ref{Systeme} describes the environment and setup parameters of the experiment.
In Section \ref{Systemi}, we present the system implementation.
The performance of the proposed system and the experimental results are demonstrated in Section \ref{Performance}.

\subsection{System environment and configuration}
\label{Systeme}
The experiments in this paper are mainly conducted in a single machine environment, and the configuration of the system is as follows.

\subsubsection{Network architecture}
The scheme consists of a total of eight network nodes, as shown in Tab. \ref{tab: Num}.

\begin{table}[htbp]
	\centering
	\caption{Number of network nodes}
	\label{tab: Num}
	\resizebox{\linewidth}{!}{
		\begin{tabular}{l|l|l}
			\hline
			\hline
			\textbf{Node name} & \textbf{Discription}       & \textbf{Number} \\
			\hline
			Fabric-couchdb     & Database node              & 4               \\
			Fabric-ca          & CA node                    & 2               \\
			Fabric-peer        & peer node                  & 4               \\
			Fabric-orderer     & orderer node               & 1               \\
			Fabric-tools       & cli node                   & 1               \\
			PSC                & policy smart contract node & 4               \\
			ASC                & access smart contract node & 4               \\
			\hline
			\bottomrule
	\end{tabular}}
\end{table}

\subsubsection{Chaincode deployment}
In Fabric-SCF, all the attribute delimitation, access control, and system operations are implemented by chaincode.
Chaincode setup mainly includes installation, instantiation, and upgrade

i): Installation: After the initialization of the blockchain network is completed, the installation of the chaincode can begin. The installation of chaincode is carried out through Hyperledger client nodes, which are used to install the chaincode into each peer node in turn.

ii): Instantiation: After the installation of the chaincode is completed, any peer node is designated to instantiate the installed chaincode.

iii): Update: Before the chaincode can be updated, a new chaincode must be installed, i.e. the chaincode update is only valid on the peer node where the new chaincode is installed.

\subsection{System implementation}
\label{Systemi}
This section mainly demonstrates the operational effects of three smart contracts to make a clearer understanding of this program, and the specific operation results are as follows.

\subsubsection{User management contract}
As section \ref{secUser Management Contract} presents, this part of the method focuses on the creation, query, and deletion of each participant $SP_i$, $Dt_j$, $CE$ in the system.

\begin{figure}[htbp]
	\centering
	\includegraphics[width=\linewidth]{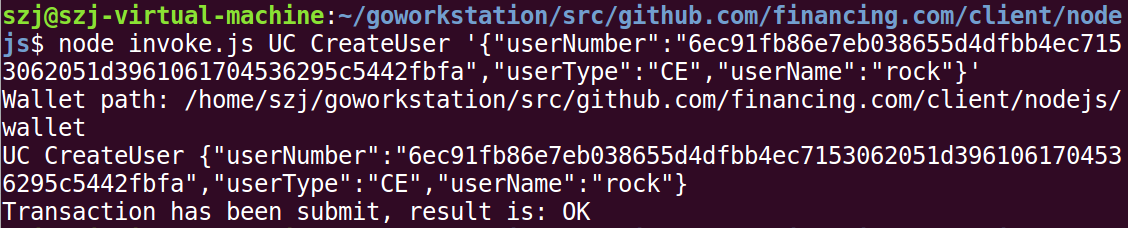}
	\caption{System response by calling the $CreateUser()$ method.}
	\label{fig:createuser}
\end{figure}

$CreateUser()$: This method can be called if you want to add a user, this method automatically generates a user number based on the user name and type, as shown in Fig. \ref{fig:createuser}.

\begin{figure}[htbp]
	\centering
	\includegraphics[width=\linewidth]{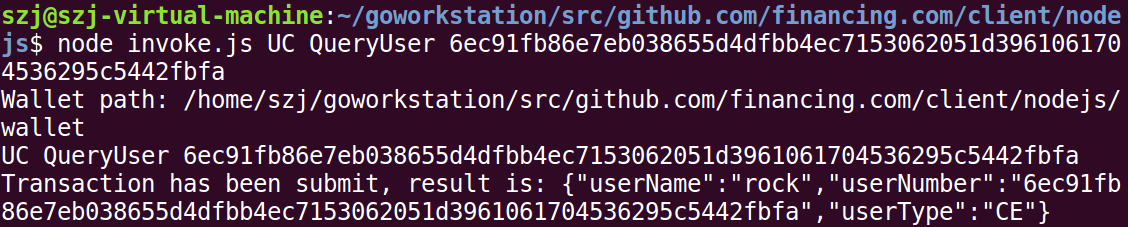}
	\caption{System response by calling the $QueryUser()$ method.}
	\label{fig:queryuser}
\end{figure}

$QueryUser()$: If you want to query the details of a user, you can call this method to query the user's information based on the user number as shown in Fig. \ref{fig:queryuser}.

\begin{figure}[htbp]
	\centering
	\includegraphics[width=\linewidth]{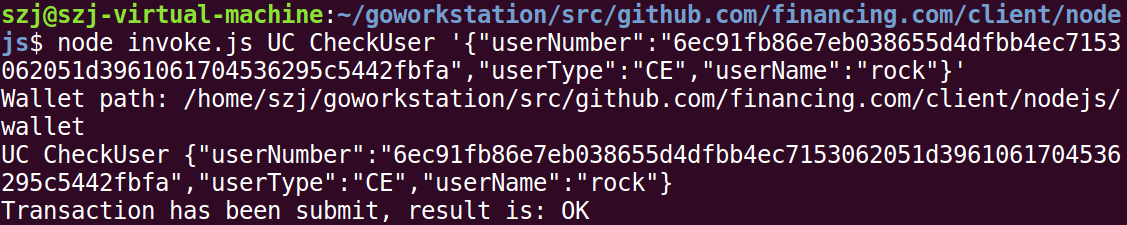}
	\caption{System response by calling the $CheckUser()$ method.}
	\label{fig:checkuser}
\end{figure}

$CheckUser()$: This method is used to check the user's identity, including his certificate, private key and identity type, etc. It returns OK if the check passes, otherwise, it returns Error as shown in Fig. \ref{fig:checkuser}.

\subsubsection{Financing project management contract}
As introduced in section \ref{secFinancing}, this part of the chaincode mainly realizes the functions of adding, querying, updating, and deleting financial projects.

\begin{figure}[htbp]
	\centering
	\includegraphics[width=\linewidth]{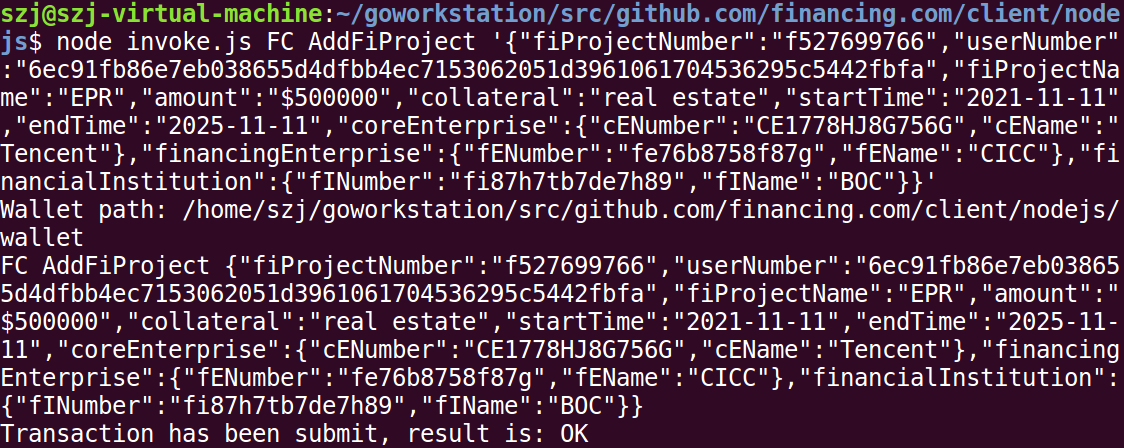}
	\caption{System response by calling the $AddFiProject()$ method.}
	\label{fig:addfiproject}
\end{figure}

$AddFiProject()$: This method can be called to write project information to the SDB when project information needs to be added, as shown in Fig. \ref{fig:addfiproject}.

\begin{figure}[htbp]
	\centering
	\includegraphics[width=\linewidth]{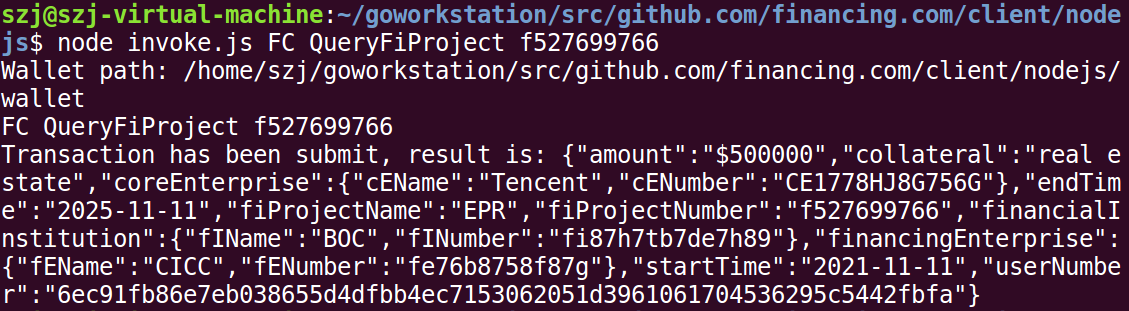}
	\caption{System response by calling the $QueryFiProject()$ method.}
	\label{fig:queryfiproject}
\end{figure}

$QueryFiProject()$: This method can be called to view information about a particular project by its project number, thus enabling the retrieval of detailed information about a financial project, as shown in Fig. \ref{fig:queryfiproject}.

\begin{figure}[htbp]
	\centering
	\includegraphics[width=\linewidth]{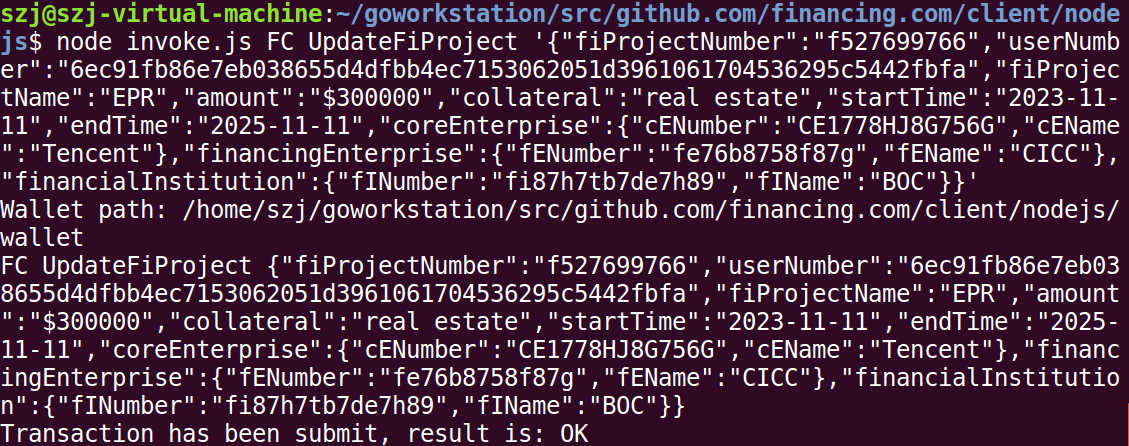}
	\caption{System response by calling the $UpdateFiProject()$ method.}
	\label{fig:updatefiproject}
\end{figure}

$UpdateFiProject()$: This method can be called when a project needs to be updated to modify the information of the financial project, before executing this method you need to execute the $CheckUdvInfo()$ method to determine if you have the update permission, as shown in Fig. \ref{fig:updatefiproject}.

\begin{figure}[htbp]
	\centering
	\includegraphics[width=\linewidth]{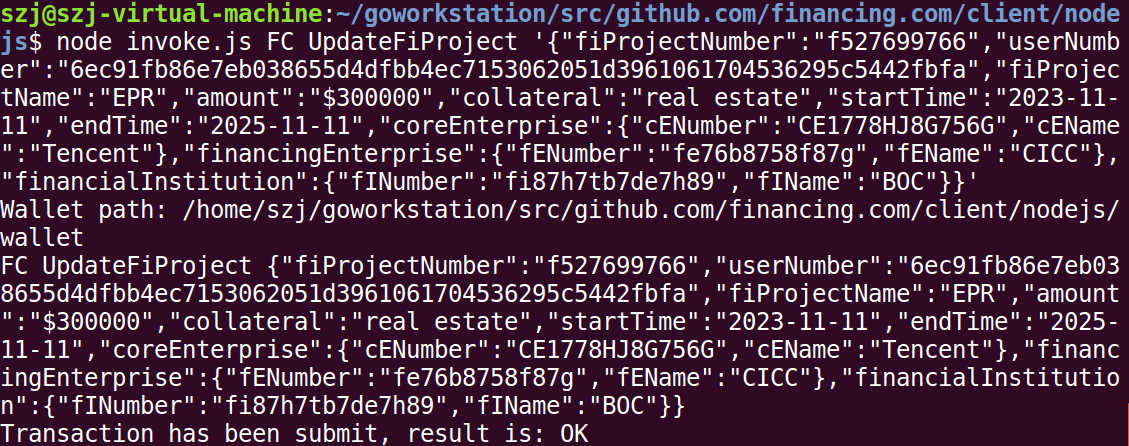}
	\caption{System response by calling the $DeleteFiProject()$ method.}
	\label{fig:deletefiproject}
\end{figure}

$DeleteFiProject()$: This method can be executed if the system needs to delete project information from the SDB as shown in Fig. \ref{fig:deletefiproject}.

\begin{figure}[htbp]
	\centering
	\includegraphics[width=\linewidth]{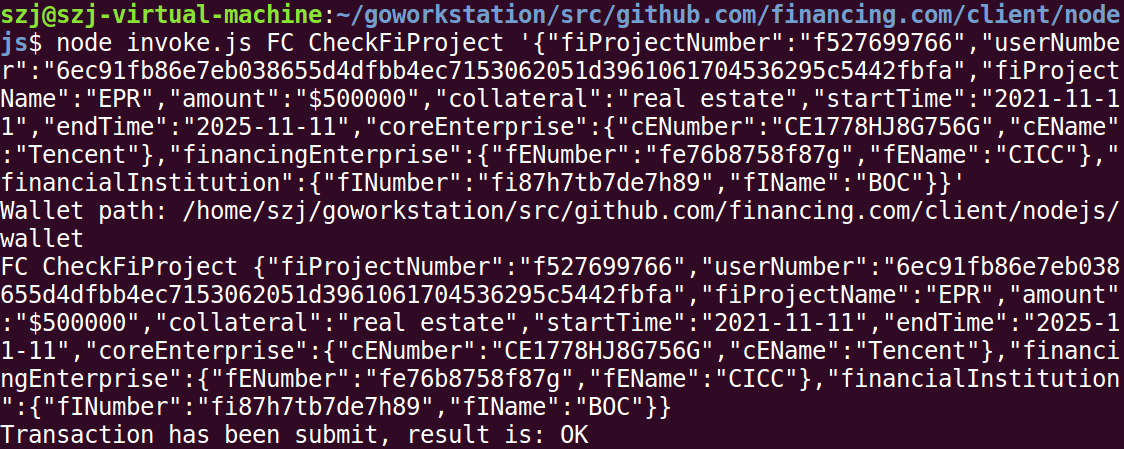}
	\caption{System response by calling the $CheckFiProject()$ method.}
	\label{fig:checkfiproject}
\end{figure}

$CheckFiProject()$:This method is used to verify the legitimacy of a financial project and to write a complete project message to the SDB once it has been verified, as shown in  Fig. \ref{fig:checkfiproject}.

\subsubsection{Access control contract}
As section \ref{AccessC} presents, this part of the chaincode is mainly used to implement auditable access control for participants.

\begin{figure}[htbp]
	\centering
	\includegraphics[width=\linewidth]{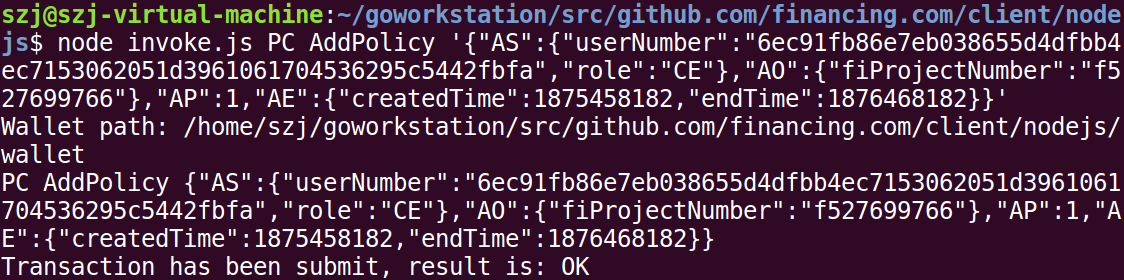}
	\caption{System response by calling the $AddPolicy()$ method.}
	\label{fig:addpolicy}
\end{figure}

$AddPolicy()$: This method is used to store the specified policy in the status database after it has been agreed between the administrator and the user as shown in Fig. \ref{fig:addpolicy}.

\begin{figure}[htbp]
	\centering
	\includegraphics[width=\linewidth]{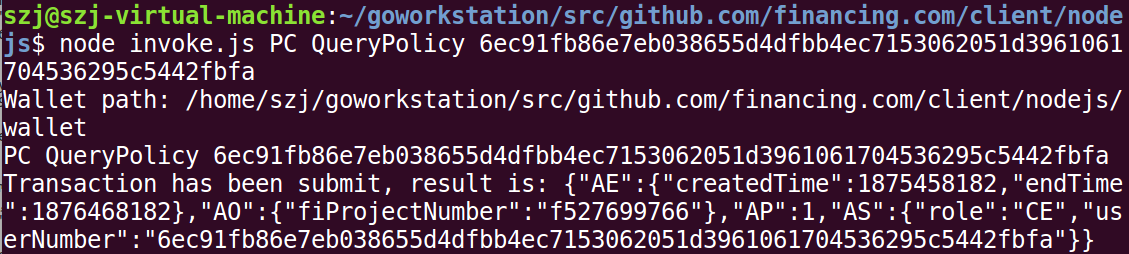}
	\caption{System response by calling the $QueryPolicy()$ method.}
	\label{fig:querypolicy}
\end{figure}

$QueryPolicy()$: This method can be used to determine if the user has access rights by querying by user number as shown in Fig. \ref{fig:querypolicy}.

\begin{figure}[htbp]
	\centering
	\includegraphics[width=\linewidth]{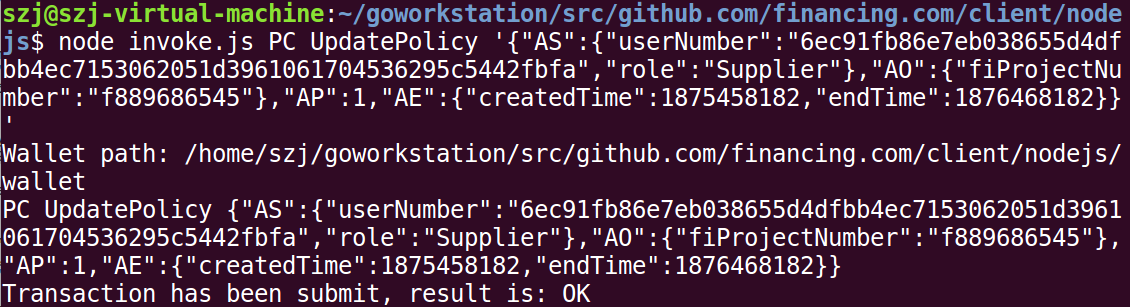}
	\caption{System response by calling the $UpdatePolicy()$ method.}
	\label{fig:updatepolicy}
\end{figure}

$UpdatePolicy()$: This method can be called to update the policy once the access requirements have changed as shown in Fig. \ref{fig:updatepolicy}.

\begin{figure}[htbp]
	\centering
	\includegraphics[width=\linewidth]{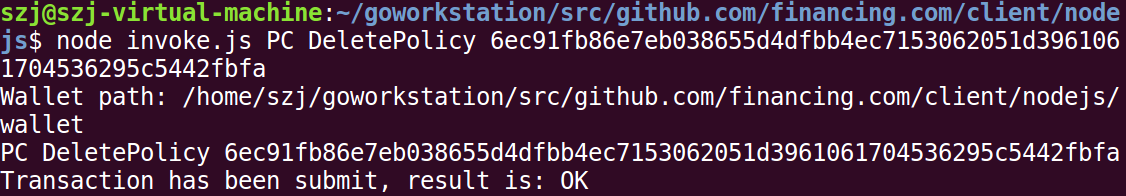}
	\caption{System response by calling the $DeletePolicy()$ method.}
	\label{fig:deletepolicy}
\end{figure}

$DeletePolicy()$: This method can be called to remove a policy if it is no longer valid or if you need to force a policy to be removed as shown in Fig. \ref{fig:deletepolicy}.

\begin{figure}[htbp]
	\centering
	\includegraphics[width=\linewidth]{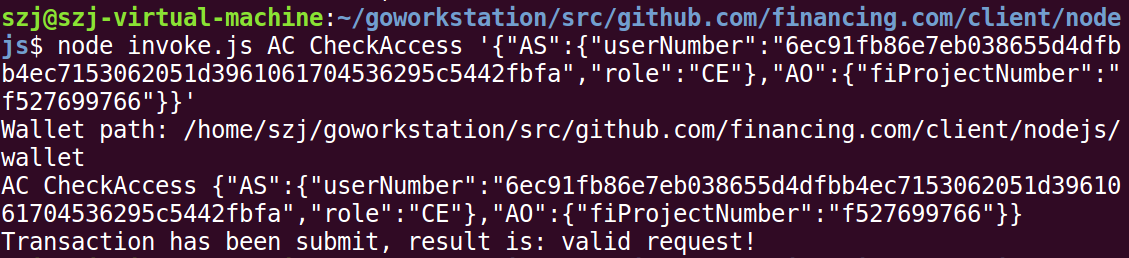}
	\caption{System response by calling the $CheckAccess()$ method.}
	\label{fig:checkaccess}
\end{figure}

$CheckAccess()$: After receiving a request from the user, the $CheckAccess()$ method in AC will be called automatically to verify whether the request is valid or not as shown in Fig. \ref{fig:checkaccess}.

\subsection{Performance tests and experimental results}
\label{Performance}
To validate the performance of the proposed Fabric-SCF, we designed two sets of simulation experiments to simulate the concurrency, transaction time, and throughput of the system under real-life usage conditions.

\subsubsection{Execution time of the contract}
\begin{figure*}[htbp]
	\centering
	\subfigure[User management contract]{
		\label{fig:usertime}
		\includegraphics[width=0.45\linewidth]{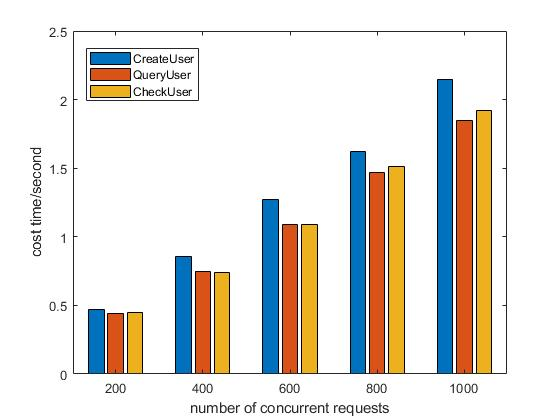}
	}
	\hspace{0.01\linewidth}
	\subfigure[Financing project management contract]{
		\label{fig:projecttime}
		\includegraphics[width=0.45\linewidth]{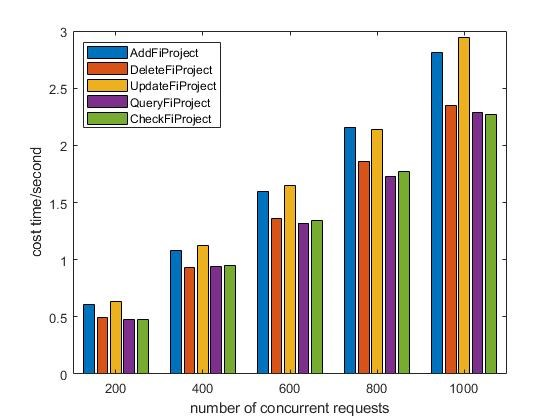}
	}
	\vfill
	\subfigure[Access control contract]{
		\label{fig:policytime}
		\includegraphics[width=0.45\linewidth]{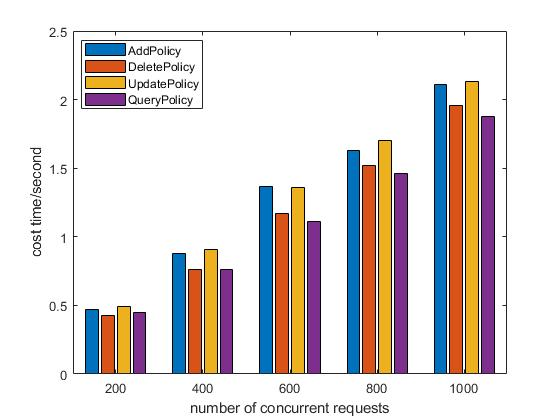}
	}
	\hspace{0.01\linewidth}
	\subfigure[Check access contract]{
		\label{fig:checkaccesstime}
		\includegraphics[width=0.45\linewidth]{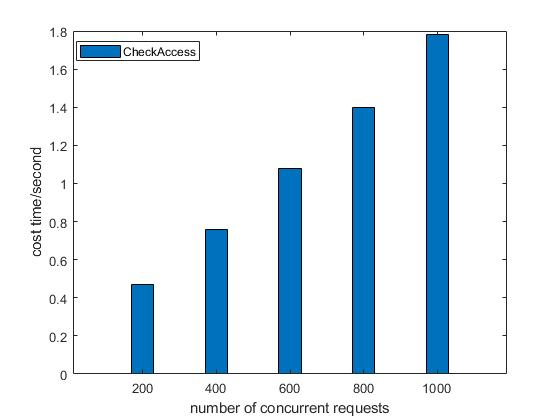}}
	\caption{Execution time under different block sizes.}
	\label{etime}
\end{figure*}

Following the system usage logic, we first tested the system response time of the user management contract, the financing project management contract, and the access control contract separately by module, as shown in Fig. \ref{etime}.
The experimental results show that with the increase of block size, the contract execution time shows an orderly growth trend and the growth is controllable.
Obviously, the execution time for $add()$ and $update()$ contracts are higher than those for $query()$ and $delete()$ contracts due to the data write operations involved, however the difference is small and reasonably manageable.

\subsubsection{TPS of the contract}
\begin{figure*}[htbp]
	\centering
	\subfigure[User management contract]{
		\label{fig:usertps}
		\includegraphics[width=0.45\linewidth]{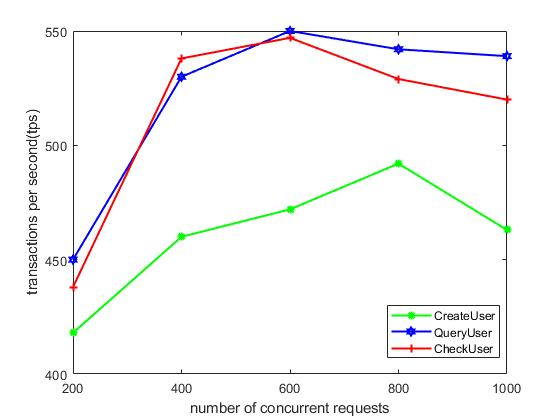}
	}
	\hspace{0.01\linewidth}
	\subfigure[Financing project management contract]{
		\label{fig:projecttps}
		\includegraphics[width=0.45\linewidth]{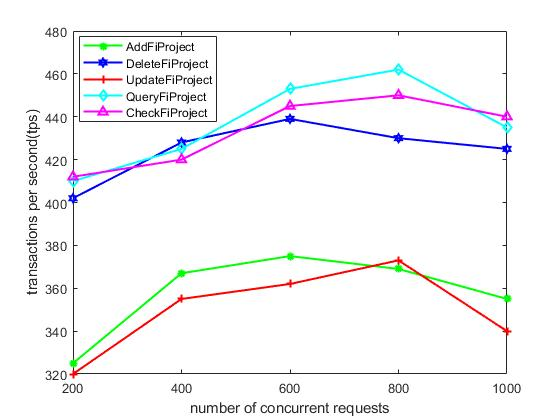}
	}
	\vfill
	\subfigure[Access control contract]{
		\label{fig:policytps}
		\includegraphics[width=0.45\linewidth]{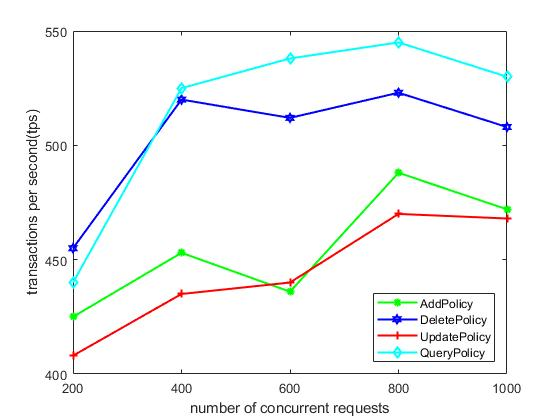}
	}
	\hspace{0.01\linewidth}
	\subfigure[Check access contract]{
		\label{fig:checkaccesstps}
		\includegraphics[width=0.45\linewidth]{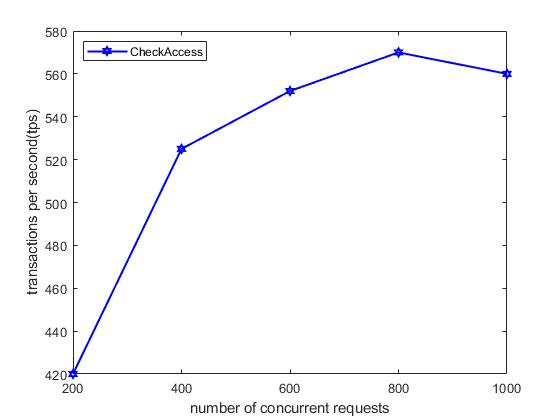}}
	\caption{Execution time under different block sizes.}
	\label{etps}
\end{figure*}

In the second group of experiments, we compared the TPS of Fabric-SCF under different concurrent numbers under simulated real conditions. 
Under the Solo and Kafka consensus mechanism, the throughput of the proposed system can be maintained at a relatively smooth level, and none of them exceeds the threshold of 550 as shown in Fig. \ref{etps}.
In conclusion, the Fabric-SCF system maintains high TPS with relatively consistent performance output under intensive requests.

\section{Conclusion and Further Work}
\label{sec:Conclusion and Further Work}
This paper proposes a blockchain-based SCF management system to enable secure data storage and auditable access control.
To accommodate the changing SCF business logic and increasing privacy requirements, the proposed system enables cross-organizational data management and access, while ensuring secure storage, efficient access, and access traceability of privacy-sensitive data.
With the combination of consensus mechanisms and smart contracts, financial project data in SCF can be securely stored in the blockchain.
Meanwhile, the access policy of the proposed system supports programmability and can therefore be modified according to different business processes.
Furthermore, thanks to the ABAC access control mechanism, participating parties can have auditable access provided they are fully authorized.
Finally, simulation experiments confirm that Fabric-SCF can guarantee system stability and TPS at larger business volumes.

Future works can consider the following aspects : 

\begin{enumerate}
	\item  The business logic of SCF is relatively simple. In the future, we consider trying to introduce more complex business processes to test the universality of the system in the financial field.
	\item Consider using large-scale test environments such as changing stand-alone environments to multi-machine environments, to simulate use scenarios more realistically. 
	\item Future works can consider combining the data storage on the chain with the data storage mode under the chain.
	\item In the future, researchers can consider deploying on a wider range of platforms, such as Ethereum, to test throughput and execution time together with gas consumption. 
\end{enumerate}
%
%

\section*{Funding}
This research is supported by the National Natural Science Foundation of China under Grant 61873160, Grant 61672338, and the Natural Science Foundation of Shanghai under Grant 21ZR1426500.

\section*{Declaration of interests}
The authors declare that they have no known competing financial interests or personal relationships that could have appeared to influence the work reported in this paper.

\ifCLASSOPTIONcaptionsoff
\newpage
\fi

\bibliographystyle{IEEEtran}
\bibliography{Ref}

\begin{IEEEbiography}[{\includegraphics[width=1in,height=1.25in,clip,keepaspectratio]{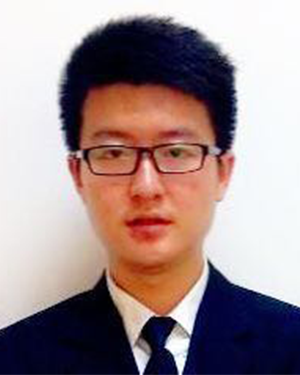}}]{Dun Li} received the B.S. degree in Human Resource Management from the Huaqiao University, Quanzhou, China, in 2013,  and the M.S. degree in Finance from the Macau University of Science and Technology, Macau, China, in 2015. He is currently doing his Ph.D. degree in Information Management and Information Systems at Shanghai Maritime University. His main research interests include smart finance, big data, machine learning, IoT, and blockchain.
\end{IEEEbiography}

\begin{IEEEbiography} [{\includegraphics[width=1in,height=1.25in,clip,keepaspectratio]{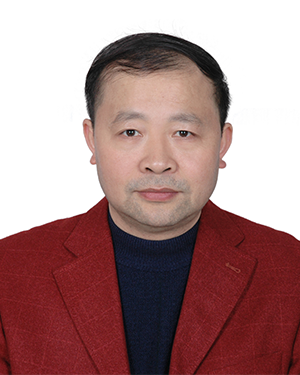}}] {Dezhi Han}
	received the BS degree from Hefei University of Technology, Hefei, China, the MS  degree and PhD degree from Huazhong University of Science and Technology, Wuhan, China. He is currently a professor of computer science and engineering at Shanghai Maritime University. His specific interests include storage architecture, blockchain technology, cloud computing security and cloud storage security technology. 
\end{IEEEbiography}

\begin{IEEEbiography} [{\includegraphics[width=1in,height=1.25in,clip,keepaspectratio]{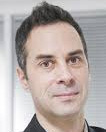}}] {Noel Crespi}
	received the master’s degrees from the University of Orsay (Paris 11) and the University of Kent, U.K., the Diplome d’Ingenieur degree from Telecom ParisTech, and the Ph.D. and Habilitation degrees from Paris VI University (Paris-Sorbonne). Since 1993, he has been with CLIP, Bouygues Telecom, and then with Orange Labs in 1995. He took leading roles in the creation of new services with the successful conception and launch of Orange prepaid service, and in standardization (from rapporteurship of IN standard to coordination of all mobile standards activities for Orange). In 1999, he joined Nortel Networks as a Telephony Program Manager, architecting core network products for the EMEA region. In 2002, he joined the Institut Mines-Telecom and is currently a Professor and the Program Director, leading the Service Architecture Laboratory. He coordinates the standardization activities for the Institut MinesTelecom, ITU-T, ETSI, and 3GPP. He is an Adjunct Professor with KAIST, an Affiliate Professor with Concordia University, and a Guest Researcher with the University of Goettingen. He is also the Scientific Director of the French-Korean Laboratory ILLUMINE. His current research interests are in sofwarization, data analysis, and Internet of Things/services
\end{IEEEbiography}

\begin{IEEEbiography} [{\includegraphics[width=1in,height=1.25in,clip,keepaspectratio]{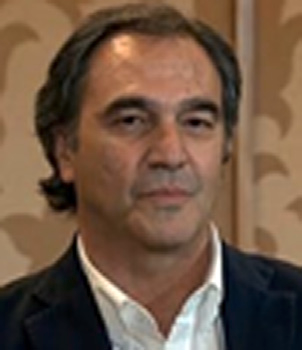}}] {Roberto Minerva}
is Associated Professor within the Service Architecture Lab of Institut Mines Telecom–Telecom Sud Paris part of Institute Polytechnique de Paris. He received the M.S. degree in computer science from Bari University, Italy and doctoral degree in Computer Science and Telecommunications from the Pierre and Marie Curie University. From 1987 to 1996, he was a researcher in the Service Architectures and Network Intelligence area within Telecom Italia Research Center. In the following years, he was appointed responsible of several research groups related to Network Intelligence evolution. From 2013 up to 2016 he was in the Strategic Initiatives of TIM. From 2016 he was the Technical Project Leader of SoftFIRE, a European Project devoted to the experimentation of NFV, SDN and edge computing. He is author of >70 article in journals and international conferences. Dr. Minerva has been the chairperson of the IEEE IoT Initiative in the period 2014–2016.
\end{IEEEbiography}

\begin{IEEEbiography}[{\includegraphics[width=1in,height=1.25in,clip,keepaspectratio]{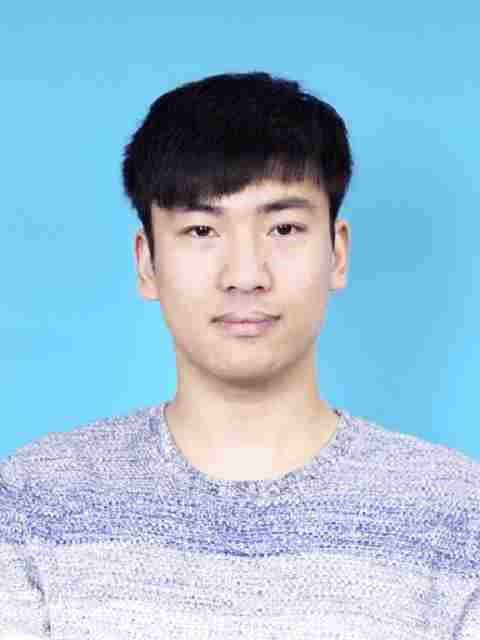}}]{Zhijie Sun} received the B.S degree in information management and information system from Henan Polytechnic University. and he is currently pursuing the M.S. degree in Shanghai Maritime University. His main research interests include blockchain technology and its applications and cryptography.
\end{IEEEbiography}

\end{document}